# Attosecond transient grating spectroscopy with near-infrared grating pulses and an extreme ultraviolet diffracted probe

Rafael Quintero-Bermudez[1,2,3], Lorenz Drescher[1,2,3], Vincent Eggers[1], Kevin Gulu Xiong[1], Stephen R. Leone[1,2,3]

[1]*Department of Chemistry, University of California, Berkeley, California 94720, USA*
[2]*Chemical Sciences Division, Lawrence Berkeley National Laboratory, Berkeley, California 94720, USA*
[3]*Department of Physics, University of California, Berkeley, California 94720, USA*

Transient grating spectroscopy has become a mainstay among metal and semiconductor characterization techniques. Here we extend the technique towards the shortest achievable timescales by using tabletop high-harmonic generation of attosecond extreme ultraviolet (XUV) pulses that diffract from transient gratings generated with 500-1000 nm sub-5 fs near-infrared (NIR) pulses. We demonstrate the power of attosecond transient grating spectroscopy (ATGS) by investigating the ultrafast photoexcited dynamics in an Sb semimetal thin film. ATGS provides an element-specific, background-free signal, unfettered by spectral congestion, in contrast to transient absorption spectroscopy. With the ATGS measurements in Sb polycrystalline thin films, we observe the generation of coherent phonons and investigate the lattice and carrier dynamics. Among the latter processes, we extract carrier thermalization, hot carrier cooling, and electron-hole recombination, which are on the order of 20 fs, 50 fs, and 2 ps, timescales, respectively. Furthermore, simultaneous collection of transient absorption and transient grating data allows us to extract the total complex dielectric constant in the sample dynamics, including the real-valued refractive index, from which we are also able to investigate carrier-phonon interactions and longer-lived phonon dynamics. The outlined experimental technique expands the capabilities of transient grating spectroscopy and attosecond spectroscopies by providing a wealth of information concerning carrier and lattice dynamics with an element-selective technique, at the shortest achievable timescales.

## INTRODUCTION

Ultrafast spectroscopy has made it possible to observe a wide range of physical phenomena unfold. Transient grating spectroscopy (TGS), in particular, has become popular among ultrafast techniques due to its versatility, sensitivity, and wealth of information available in the data. Among the latter, TGS is capable of quantifying diffusion lengths of particles and quasi-particles, such as charge carriers, spins, and phonons (*1*). Additionally, the introduction of an additional reference beam makes TGS uniquely suited to probe both refractive index as well as absorption dynamics (*2*).

TGS is a subset of four-wave mixing nonlinear spectroscopy in which a probe beam is used to measure the dynamics of a time-dependent grating. In a typical TGS measurement scheme, two non-collinear pump pulses are employed to generate a pump grating on the sample. A temporally delayed probe pulse is diffracted by the resulting *amplitude* and *phase* gratings. These two components result from absorptive and dispersive phenomena, or more concretely from changes caused by the pump pulses in the attenuation coefficient and refractive index, respectively. Unlike transient absorption spectroscopy (TAS) and transient reflectivity spectroscopy (TRS) measurements, which are used to track dynamics predominantly in the absorption and refractive index, respectively, the TGS signal consists of both terms (*2*):

$$\eta_{TOTAL} = \eta_{AMP} + \eta_{PHA} = \left[\left(\frac{\Delta k d}{4}\right)^2 + \left(\frac{\pi \Delta n d}{\lambda}\right)^2\right] e^{-\alpha d},$$

where $\eta_{TOTAL}$ is the total grating efficiency, $\eta_{AMP}$ and $\eta_{PHA}$ are the amplitude and phase grating efficiencies, $\Delta k$ is the change in the absorption, $\Delta n$ is the change in refractive index, d is the sample thickness, $\lambda$ is the probe beam wavelength, and $\alpha$ is the sample attenuation coefficient.

TGS has been used to investigate a wide range of electrical, thermal, and mechanical properties and phenomena in condensed matter systems (*1*, *3*), including photocarrier dynamics such as electrons and holes in metals and semiconductors (*4*, *5*), excitons in semiconductors (*6*, *7*), and quasi-particles in superconductors (*8*); phonon dynamics (*9*, *10*); spin dynamics (*11*); magnetization dynamics (*12*); charge density waves (*13*); lattice strain (*14*); and surface acoustic waves (*15*). Recent studies at Free Electron Laser (FEL) facilities have shown that extreme ultraviolet (XUV) pump – XUV probe TGS could be used to investigate nanoscale charge-carrier diffusion due to the nanoscale grating period (*16–19*). However, the temporal resolution of the measurement was limited to 40-70 fs due to the pulse width of the FEL source. TGS signals were also demonstrated in tabletop XUV systems, however these were limited to resolutions on the order of a few 100 fs (*20*).

Attosecond studies of condensed matter systems have elucidated a number of phenomena including coherent phonons (*21*, *22*), carrier thermalization (*23*), and core excitons (*24–26*), due to the temporal resolution (*24*), element-selective detection (*27*), and ability to isolate multiple processes obtained within their large spectral bandwidth datasets, such as separate extraction of electron and hole-phenomena (*28*). These, however, have been mostly limited to TAS. Expanding attosecond methodologies in condensed matter systems to nonlinear and non-collinear measurement schemes could lead to enhanced signal-to-noise, background-free data collection that is free of spectral congestion, and richer datasets with information beyond the limits of TAS.

Here, we report the first TGS measurements capable of achieving attosecond resolution (ATGS). We use tabletop high-harmonic generation (HHG) of XUV pulses (*27*) to generate attosecond probe pulses that diffract from transient gratings generated with sub-5 fs near infrared (NIR) pulses. We benchmark the method by investigating the ultrafast photoexcited dynamics in Sb semimetal thin film, which exhibits coherent phonon oscillations. With the technique we are able to resolve carrier thermalization and hot carrier cooling, which would be inaccessible by fs TGS techniques. We also observe electron-hole recombination, carrier-phonon interaction, and longer-lived phonon dynamics. We then use simultaneous collection of attosecond transient absorption spectroscopy (ATAS) and ATGS to determine ultrafast changes in the refractive index, providing a more complete picture of the photoexcited dynamics. We discuss the potential of ATGS for condensed-matter systems in light of these exciting results.

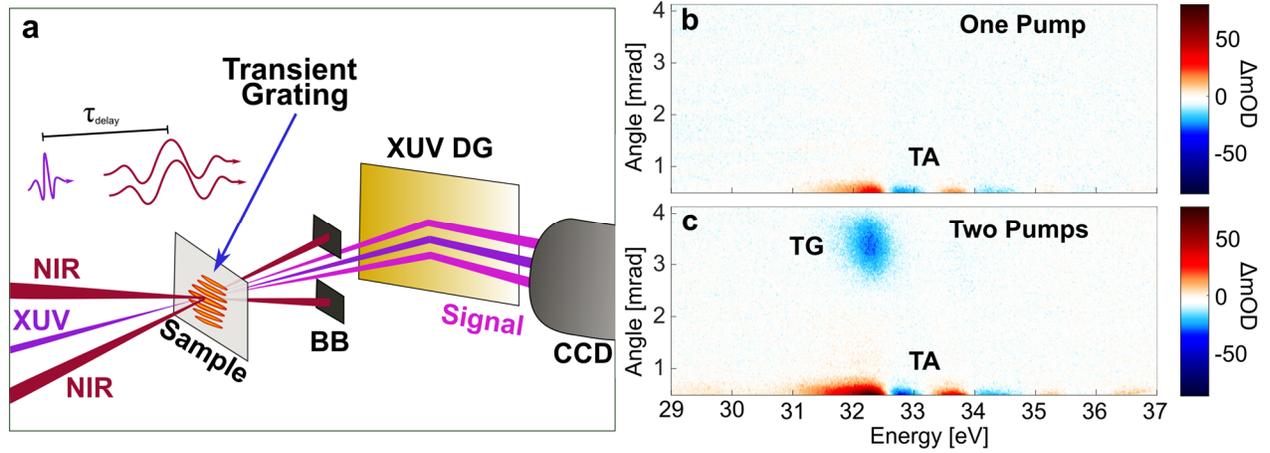

Figure 1. **Attosecond transient grating experiment.** **a** Schematic of experimental setup consisting of two non-collinear NIR few-cycle pump pulses and one time-delayed attosecond XUV probe pulse. A transient grating consisting of amplitude and phase terms is generated at the sample, which produces the off-axis diffraction of the XUV probe pulse. These diffracted beams are the detectable signal pulses, which are collected with a CCD after being spectrally dispersed with an XUV diffraction grating. BB – beam block, CCD – charge-coupled device camera, DG – dispersion grating. **b** Differential camera image at 150 fs time delay between pump and probe pulses with one pump arm only and **c** with both pump arms, demonstrating the experimental principle of the grating diffraction in a thin film Sb sample. The camera has been shifted such that transient absorption for the collinear XUV probe beam can be just observed simultaneously with the off-axis transient grating signal.

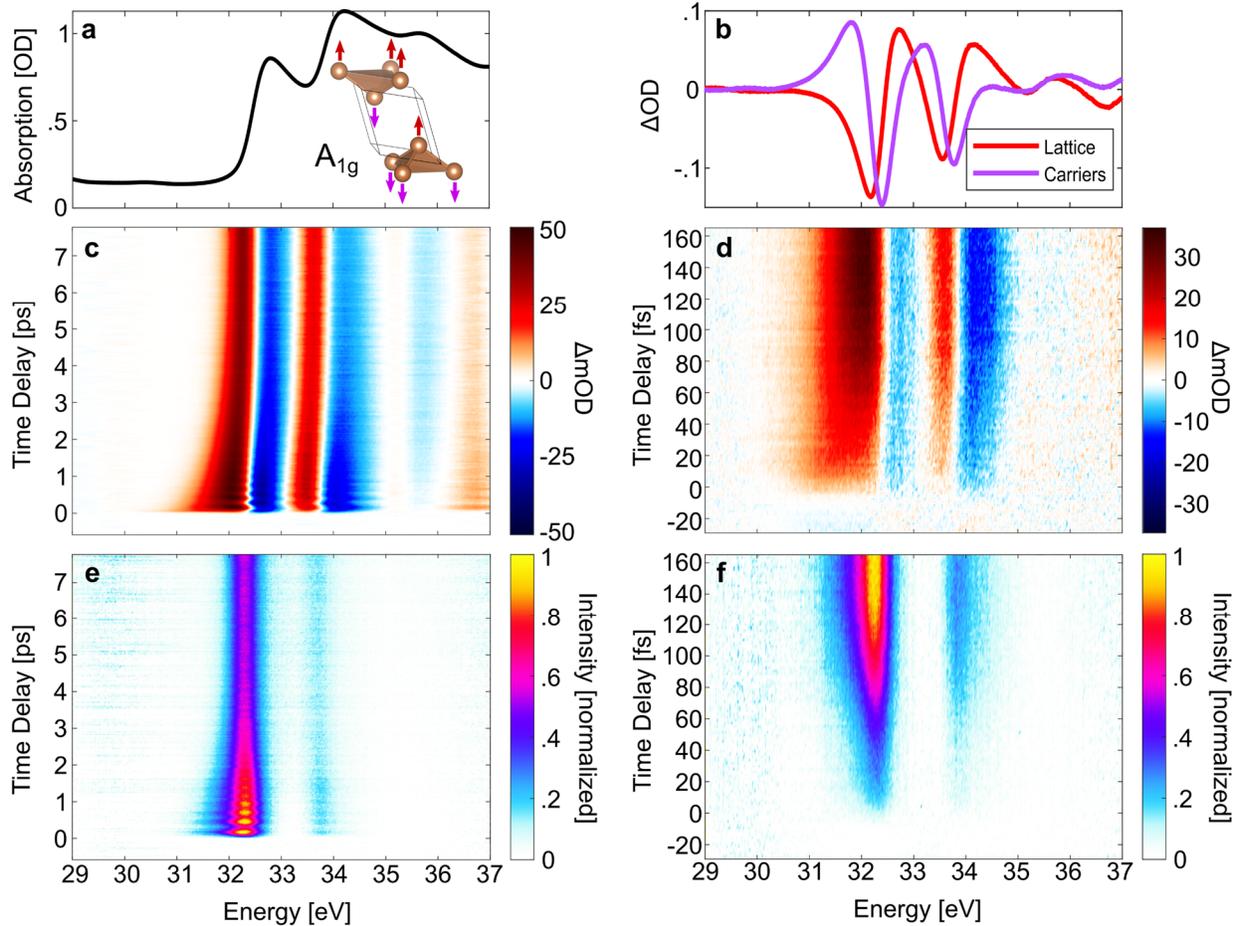

Figure 2. **Simultaneous collection of XUV transient absorption and transient grating spectroscopy data in Sb.** **a** XUV linear absorption spectrum in Sb at $N_{4,5}$ edges. Inset: $A_{1g}$ phonon mode is depicted on the Sb crystal structure. **b** Change in absorbance due to lattice and carrier dynamics as obtained from singular value decomposition of transient absorption data. **c** ATAS data in Sb reveals the excitation and decay of coherent phonon dynamics that decay over the course of a few ps. **d** ATAS data collected with a higher temporal resolution during the signal rise can be used to identify hot carrier relaxation. **e** ATGS data in contrast depicts a simplified spectrum that together with ATAS provides a richer picture of the photoexcited dynamics in solid-state materials. **f** ATGS data collected with a higher temporal resolution during signal rise provides analogous information to **d**.

## RESULTS

The experimental methodology is illustrated in Fig. 1a. A detailed account of the methodology is outlined in a later section. In brief, two coincident non-collinear sub-5 fs 500-1000 nm NIR pump pulses are incident on samples followed by a 25-45 eV attosecond XUV probe pulse that is generated by HHG, in which a portion of the aforementioned NIR pulse energy is incident onto a Kr gas cell. The XUV beam is spectrally dispersed by a grating and recorded with an X-ray camera. The TGS measurement results in two non-collinear signal beams that do not overlap with the pump or probe beams (Fig 1b). It is possible to simultaneously detect the signal and probe beams with the XUV charge-coupled device camera (CCD). As a result, it is possible to gather spectrally broad ATAS and ATGS data at the same time.

The experimental technique is applied to measurements on an Sb semimetal thin film with a NIR pump intensity of 7.9 mJ/cm$^2$. A NIR-XUV offset angle of 36 mrad is used, which yields a grating period of approximately 11 μm ($\Lambda = \frac{hc}{E_{IR}\sin\theta}$, where $E_{IR}$ is the NIR pump photon energy and θ the pump beam offset angle). Samples consist of 30 nm thick Sb thin films evaporated onto 30 nm thick $Si_3N_4$ membrane substrates. The XUV linear absorption spectrum is depicted in Fig 2a. Sb is known to exhibit a coherent phonon motion due to the lifting of a Peierls distortion upon ultrafast photoexcitation (21, 29–31). This phenomenon is attributed to displacive excitation of coherent phonons (DECP). The $A_{1g}$ phonon vibration that is active during DECP is illustrated in the inset of Fig 2a (32, 33). Upon photoexcitation, ATAS reveals the emergence of coherent phonon oscillations that decay shortly after 1 ps as well as longer-lived exponentially decaying features (Fig 2c). Using singular value decomposition (SVD) to facilitate analysis of the transient data, two main components are identified, which are attributed to lattice and carriers as done in previous work (21). The spectral shape of these components is depicted in Fig 2b. A higher temporal resolution ATAS scan is also performed during the first 160 fs to facilitate the identification of shorter-lived processes (Fig 2d).

As mentioned earlier, ATGS data are collected simultaneously during the ATAS measurements shown in Fig 2c and 2d. These are depicted in Fig 2e and 2f. While the ATAS data depict a convoluted array of spectrally-congested overlapping features due to transitions from the spin-orbit split $4d_{5/2}$ and $4d_{3/2}$ core levels, the ATGS can be more easily interpreted since the ATGS consists only of positive signal and both spin-orbit components are clear and well-separated.

The ATAS data is first analyzed by fitting the lattice and carrier transient data to a DECP model outlining the decay of the coherent phonon oscillations, carrier relaxation, and carrier recombination (*21*, *29*). The first and second SVD components of the ATAS, $C_1$ and $C_2$ (for the data of Fig 2c), which correspond to lattice and carrier dynamics respectively (*21*), are plotted in Fig 3a and Fig 3b respectively, along with the optimized fit to Equations SE1 and SE2. The optimal fit parameters are included in Table ST1. The fit parameters are comparable to those determined in previous work (*21*, *29*, *31*), especially after considering the greater phase offset and shorter lifetimes that are expected at higher NIR pump intensities (*31*). Most notably, our measurements find an $A_{1g}$ phonon frequency of 4.13±0.02 THz, a carrier recombination lifetime of 2.07±0.04 ps, and a coherent phonon oscillation decay of 540±50 fs.

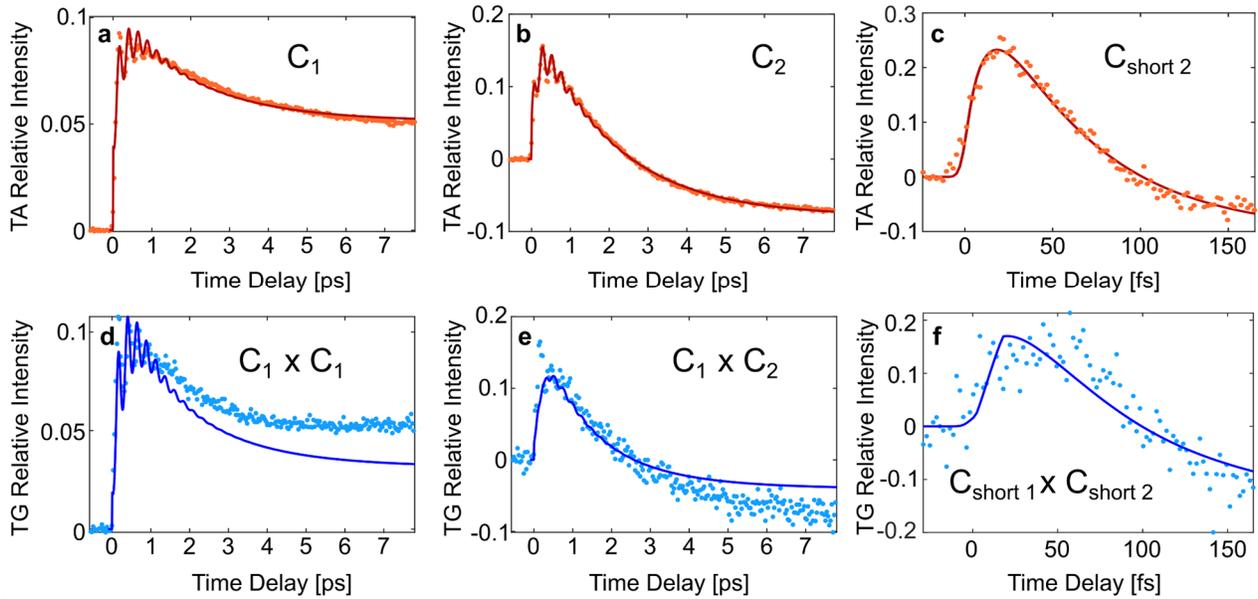

**Figure 3. Fitting transient absorption and transient grating data with Sb photoinduced dynamics model. a** Experimental ATAS data singular value decomposition (SVD) component 1 and **b** component 2, corresponding to lattice and carrier, respectively are depicted (dots) along with the fit to equations SE1 and SE2 (line). **c** Experimental ATAS data SVD component 2 for the high temporal resolution scan in Figure 2d, corresponding to hot carrier relaxation, depicted (dots) along with fit to equation SE3 (line). **d** Experimental ATGS data SVD component 1 and **e** component 2, corresponding to mixing of ATAS components 1 and 2, depicted (dots) along with the product of ATAS SVD component fits $C_1 \times C_1$ and $C_1 \times C_2$, respectively (line). **f** Experimental ATGS data SVD component 2 for high temporal resolution scan, corresponding to mixing of TA components 1 and 2, depicted (dots) along with ATAS SVD component fits $C_{short1} \times C_{short2}$ (line).

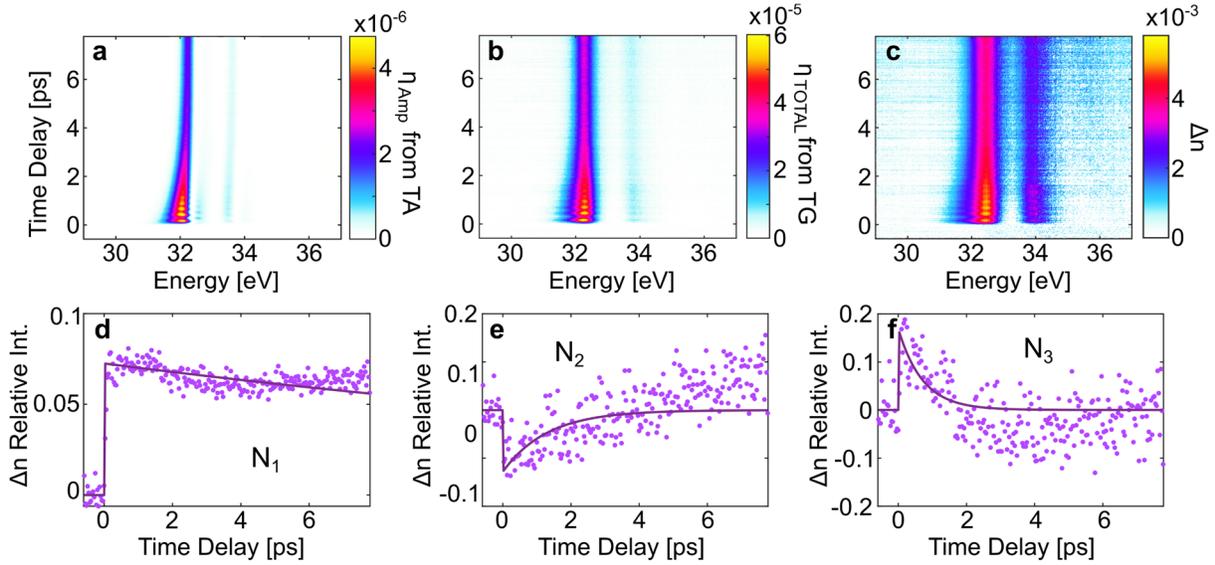

**Figure 4. Extracting transient refractive index from ATAS and ATGS.** Transient refractive index determined from ATAS and ATGS reveal the rich array of information provided by a single XUV ATGS measurement. **a** Calculated diffraction efficiency of ATGS resulting from experimentally measured ATAS data, i.e. amplitude grating efficiency. Calculated efficiency also accounts for transmittance through the sample. **b** Experimentally measured total ATGS efficiency (includes both amplitude and phase grating components). **c** The phase grating component is calculated from **a** and **b** and used to extract the transient refractive index in Sb after photoexcitation. **d** First SVD component of transient refractive index dynamics in **c** reveal the presence of long-lived phenomena in the refractive index (and ATGS) not exhibited by the ATAS, which can be attributed to phonon, carrier-phonon, and phonon-phonon dynamics. **e** Second and **f** third SVD components of transient refractive index dynamics can be attributed to lattice and carrier dynamics, respectively.

The high temporal resolution ATAS scan (Fig 2d) is used to determine carrier thermalization and relaxation. The first two SVD components of the short scan ($C_{short1}$ and $C_{short2}$) do not match those of the long scan. Instead, the first component resembles a sum of $C_1$ and $C_2$, depicting a monotonic rise in signal (Fig S4), whereas the second component depicts a delayed rise and subsequent decay (Fig 3c). The latter is too fast to be captured by the SVD analysis of the longer scan. Both the rise and decay of $C_{short2}$, it should be noted, occur with a temporal component far slower than the temporal bandwidth of the NIR pump beam. The temporal trace is fitted to the sum of an exponential rise and an exponential decay; the best fit corresponds to a 16±4 fs rise and a 45±8 fs decay. These are attributed to carrier thermalization and hot carrier cooling, respectively.

As outlined in Eq 1, transient grating efficiency is a function of the square of transient absorption modulation. As a result, the ATGS data is a function of the square of the ATAS data (*34*). The first SVD component of the ATGS data in Figure 2e is therefore found to resemble the squared $C_1$ component of the ATAS data (Fig 3d), whereas the second SVD component is found to resemble the product of the $C_1$ and $C_2$ components of the ATAS data (Fig 3e). More generally, the SVD components of the ATGS can be described by the relation: $D_i = \sum_{i,j=0}^{n} C_i C_j$, where n is the number of non-trivial SVD components identified in the ATAS, and $C_i$ and $D_i$ represent the ATAS and ATGS SVD components, respectively. In the plots, the ATGS SVD components are compared to the products of the best fit equations for ATAS SVD components: $C_1$ x $C_1$ and $C_1$ x $C_2$, respectively. The data matches the products of the fitted equations best at early time delays and diverges at later time delays; this is discussed below. The high temporal resolution ATGS scan can in turn be used to observe carrier thermalization and hot carrier cooling in its second SVD component (Fig 3f), as was done with the ATAS. In this case, the ATGS data is found to resemble the product of $C_{short1}$ x $C_{short2}$. The ATGS $C_{short2}$ is plotted alongside the product of the best fit equations for ATAS SVD components: $C_1$ x $C_{short2}$. $C_1$ was chosen as there is an analytical equation available for this component, however $C_2$ would work equally well given the nature of the monotonic rise of these components during the first 150 fs. As a result, it is found that the ATGS measurements can be used to directly analyze the ultrafast dynamics in Sb as has been found with ATAS. However, the discrepancy at greater time delays would limit its ability to interpret phenomena occurring at longer time scales.

The source of this disagreement lies in the second term of Eq 1. The ATGS signal is not simply a function of the ATAS, but instead also dependent on the photoexcited change in refractive index, $\Delta n$. The remaining discrepancy between the squared ATAS and the ATGS can thus be attributed to $\Delta n$.

Conversely, if ATAS and ATGS are both simultaneously acquired, as has been done in this work, it is possible to extract $\Delta n$, expanding the capabilities of the measurement.

To calculate $\Delta n$, the grating efficiency resulting only from the amplitude grating, $\eta_{AMP}$, is calculated from the measured ATAS (i.e., $\Delta kd$), $\eta_{AMP}(\lambda) = \left(\frac{\Delta k(\lambda) d}{4}\right)^2$. The amplitude grating efficiency is depicted in Fig 4a. The total grating efficiency (Fig 4b), $\eta_{TOTAL}$, is calculated from the XUV spectrum of the probe beam, $HHG(\lambda)$, and the raw ATGS data, $TG(\lambda)$, with the equation $\eta_{TOTAL}(\lambda) = \left(\frac{TG(\lambda)}{HHG(\lambda)}\right)$. The phase grating efficiency, $\eta_{PHA}$, and the transient change in refractive index, $\Delta n$, can then be determined by Eq 1 as: $\eta_{PHA} = \eta_{TOTAL} - \eta_{AMP}$ and $\Delta n(\lambda) = [\eta_{TOTAL}(\lambda) e^{\alpha(\lambda) d} - \eta_{AMP}(\lambda)]^{1/2} \frac{\lambda}{\pi d}$, where the sample attenuation coefficient $\alpha(\lambda)$, can be determined by performing a static absorption measurement of the sample. The experimentally determined refractive index dynamics, $\Delta n(\lambda)$ are presented in Fig 4c. The extracted $\Delta n$ for Sb exhibits some similarities to the ATGS, namely coherent phonon oscillations and some decaying components. SVD analysis is then used to determine the different components contributing to $\Delta n$. The temporal evolution of the first three SVD components $N_1$, $N_2$, and $N_3$ are plotted in Fig 4d-f, along with the best exponential decay fits. The first component, $N_1$, clearly consists of long-lived processes. The component is sufficiently well-fitted with a single exponential decay of 30 ps lifetime. However, the increasing error at larger time delays suggest that a biexponential decay may be a more accurate model. This decay lifetime cannot be estimated very accurately with such a comparatively short time scan. $N_2$ is best fitted to a decaying exponential with a 1.3±0.3 ps time constant. And $N_3$ reveals a relatively faster 720±140 fs decay. These are further discussed in the following section.

## DISCUSSION

These results show that ATGS provides an element-selective background-free technique that provides richer data with information on both dispersive and absorptive phenomena in contrast to XUV TAS. Interestingly, our measurements find that ATGS at a few-fs scale is mainly dominated by absorptive effects, whereas at longer timescales $\Delta n$ effects need to be included. As a result, ATGS can be used to accurately analyze fast electronic processes in solids, including carrier thermalization and relaxation, as well as recombination in metals, without much post-processing. However, slower processes may be more difficult to isolate due to the presence of dispersive and absorptive phenomena, as well as the manifestation of diffusion features in ATGS (1). Simultaneous collection of ATAS, however, can be used to isolate absorptive and dispersive effects, and disentangle various phenomena as we have shown.

Researchers have previously used attosecond reflectivity measurements (35, 36) to extract the ultrafast transient complex-valued dielectric function of materials after photoexcitation. In these studies, it was difficult to conclusively disentangle various carrier and phonon processes in the rich but spectrally-congested data. Additionally, XUV transient reflectivity was found to depend primarily on dispersive phenomena. However, it is known that absorptive and dispersive components in XUV spectroscopy contain complementary information. Although Kramer-Kronig's relations state that all information contained in the dispersive component must be contained in the absorptive component (37), this is only true if the entire spectrum is known. It is thus it is easier to distinguish non-absorptive phenomena, such as lattice effects, in $\Delta n$ than in the absorption spectrum (36). Additionally, in the ATGS measurement, the grating lines make it possible to quantify diffusion lifetimes for carriers, phonons, and heat, when the grating period is shorter than the mean-free path of the particle in question (1).

In the extracted $\Delta n$ refractive index information of the thin film Sb sample, $N_2$ and $N_3$ may be directly linked to lattice and carrier dynamics, respectively, as observed in the ATAS (Fig 3a,b) and ATGS (Fig 3d,e) results. $N_1$, in contrast, exhibits one or more very long-lived decaying exponentials. Given that this term is not observable in the ATAS, $N_1$ can be attributed to a diffusion process. Since the grating period used for this experiment was greater than the mean-free path of the carriers and phonons (38), it is not possible to determine the diffusion lifetimes of these. However, it is still possible to observe carrier-phonon and phonon-phonon processes.

In Bi, Lopez argued that the recombination process between the electron in the conduction band edge at the *L* point and the hole in the valence band edge at the *T* point would lead to the emission of one acoustic and two optical phonons at *X* (39). Since Sb also exhibits its conduction and valence band edges at the same points, similar phonon processes are expected (40). Additionally, further generation of acoustic phonons is expected as optical and acoustic phonons scatter. As a result, $N_1$ most likely represents carrier-phonon and phonon-phonon based heat diffusion processes.

The transient absorption collected from two beams that write a grating is also compared to transient absorption conducted with one beam in order to ensure that there are no additional grating-induced phenomena exhibited in the two-beam ATAS results. These spectra are identical (Fig S3).

We have demonstrated that the ATGS technique in solid-state materials can achieve attosecond time resolution with XUV broadband information. Attosecond resolution is necessary to observe and investigate hot electron thermalization and relaxation and can be used in future experiments to analyze ultrafast processes that are too rapid for fs experiments such as core exciton dynamics. Additionally, the broadband XUV pulses provide a rich dataset that contains element-selective information and can be processed to obtain separate information on electron and hole dynamics and diffusion in semiconductors. Furthermore, in contrast to spectrally narrow TGS experiments, ATGS probes a wide range of energies at and away from resonant transitions where amplitude and phase grating processes dominate, respectively. As a result, the data contain multispectral information on the different transient dispersive and absorptive processes and ultimately reveal different phenomena taking place in solid-state materials.

The technique presented here also demonstrates the use of simultaneous collection of ATAS as well as ATGS to extract the time-resolved change in the refractive index. This measurement provides an enhanced view of non-absorptive dynamics such as lattice and thermal dynamics. Finally, an important feature of TGS experiments lies in its ability to observe and quantify real-space dynamics, such as diffusion processes in carriers, phonons, and heat. ATGS provides access to these phenomena with unprecedented temporal resolution.

The solid-state samples investigated in this study consist of thin polycrystalline films compatible with transmission-mode XUV experiments. However, the nanometric dimensions of the average crystallite will limit the ability of TGS experiments to capture phonon diffusion lifetimes. Future ATGS experiments can be conducted in reflection mode on single crystalline samples to allow the extraction of phonon diffusion lifetimes. Charge carrier diffusion rates have been notably measured with XUV-XUV TGS experiments at FEL facilities (*16*, *19*) due to the higher flux available in such experiments that enables the use of XUV pump pulses that can generate nanoscale gratings. Future experiments should seek to combine the strengths of both experiments, i.e. the flux at FEL facilities and the consequent access to nanoscale investigations and the time resolution and spectral bandwidth in the ATGS measurements presented here, as HHG source fluxes and efficiencies are increased and as FEL temporal pulse widths are decreased. With such capabilities it could be, for instance, possible to measure charge dynamics and diffusion lengths of carriers before relaxation and before thermalization.

Future measurements could also make use of a heterodyne setup to separate amplitude and phase grating information, or absorptive and dispersive processes, to more concretely distinguish and isolate dynamics in condensed matter systems. Such experiments would require a co-propagating XUV beam that is delayed by a fixed offset in order to obtain purely amplitude and purely phase grating ATGS datasets. It should be noted, however, that this will require a delay between beams on the order of a fraction of an optical cycle, which for XUV photons will present a considerable challenge.

Finally, using cross-polarized NIR beams is suggested as a means of creating spin gratings and investigating spin dynamics and diffusion with attosecond time resolution and with the broad bandwidth inherent to ATGS. Such measurements could then be conducted on spin density waves (*41*), spin defects (*42*), as well as spin diffusion, transfer, and relaxation more generally.

## METHODS

The ATGS experiments conducted for this work use spectrally broad and temporally short pulses to evaluate a large spectral region at attosecond scales. The NIR pulses are produced by a Ti:sapphire laser (Coherent Astrella, 1 khz repetition rate, 7 mJ pulse energy, 35 fs pulse duration and 32 nm bandwidth). 3 mJ of the beam is focused with an f=2.5 m dielectric concave mirror into a 2.2 m long hollow core fiber (inner diameter 700 µm) filled with an Ar gas pressure gradient (approximately 5 Torr to 120 mTorr) to generate a supercontinuum spanning 500–1000 nm wavelengths with self-phase modulation (Fig. S2). Dispersion is compensated using eight pairs of double angled broadband chirped mirrors (Ultrafast Innovations, PC1332), fused silica wedge pairs, and potassium dihydrogen phosphate (KDP) wedge pairs and temporally compress the pulses to ~5 fs durations. A broadband 90:10 beamsplitter (Layertec) is then used to split the beam into two parts. Both beam paths after the split have a pair of fused silica wedges to independently tune the dispersion of the pump and probe beams. The larger power beam is sent into a gas cell with a 200 µm hole with Kr gas at ~15 Torr in a vacuum chamber held at ~$10^{-7}$ Torr to generate broadband XUV pulses via HHG. A 100 nm thick Al foil filter (Lebow Company) is then used to filter out the NIR and transmit the XUV pulse, which is focused onto the sample with a gold-coated toroidal mirror to act as the probe beam. The remaining NIR beam after the beamsplitter is used for the pump beams. A second broadband 50:50 beamsplitter (Layertec) is used to generate two NIR pump arms. These are then focused non-collinearly onto the sample at a θ=36 mrad angle offset with an f=1 m silver concave mirror. The XUV probe is delayed with respect to the NIR pump beams with a piezo stage (Physik Instrumente, P-620.1CD) mounted onto a stepper motor stage (Physik Instrumente, M-505-6DG). The former is used for the high temporal resolution scan while the latter is used for the longer delay scans. The pulse energy of the NIR beam is adjusted with an iris and monitored with a power meter and the beam diameter is characterized using a beam profiler (DataRay WinCamD). The NIR pump intensity was computed by accounting for the light reflected at the Sb sample surface (approximately 75% within the 500-1000 nm range (*43*)) and subtracting that from the incident fluence. We estimate an incident pump fluence of 7.5 mJ/cm$^2$.

Spatial and temporal overlap is determined by placing an Ar gas cell in the interaction region instead of the sample. Time zero is estimated by fitting the rise time of the Ar 3s3p6np autoionization signal to an exponentially-modified Gaussian function. The temporal pulse width of the NIR pump is estimated to be under 5 fs with a dispersive characterization of ultrafast pulses (D scan, Sphere Ultrafast Photonics) (Fig S1) and confirmed with the rise time of the Ar 3s3p6np autoionization signal. The spectral axis is also calibrated with the autoionization lines of Ar 3s3p6np(*44*). An additional 100 nm thick Al foil is then used to filter out any remaining NIR; meanwhile the XUV beam is spectrally dispersed by a gold-coated flat-field grating (01-0639, Hitachi) and measured with an XUV CCD camera (Pixis XO 400-B, Princeton Instruments). The XUV CCD is placed on a dual axis stage in order to shift the spectral axis and locate ATAS and ATGS, as needed.

The samples measured are 30 nm polycrystalline Sb thin films (*21*) evaporated onto 30 nm thick Si$_3$N$_4$ membranes (Norcada NX5050X). Samples are mounted onto an x,y-axis translation stage (Physik Instrumente) which is used to shift the illuminated area of the sample every 5 minutes to avoid artifacts from sample heating and damage.

The measured ATGS spectra are divided by the XUV harmonic spectral flux used during the experiments in order to normalize the ATGS and prevent inaccurate conclusions due to enhancements in the transient grating efficiency from higher XUV spectral intensities at different parts of the harmonic spectra.

## ACKNOWLEDGEMENTS

This work was supported by Air Force Office of Scientific Research (AFOSR) Grant FA9550-20-1-0334 (R.Q.B, K.G.X., and S.R.L.) and FA9550-19-1-0314 (L.D. and S.R.L.). R.Q.B. acknowledges support from Natural Sciences and Engineering Research Council of Canada (NSERC) Postdoctoral Fellowship. L.D. acknowledges the European Union's Horizon research and innovation program under the Marie Sklodowska-Curie grant agreement No. 101066334 – SR-XTRS-2DLayMat.